\documentclass[conference]{IEEEtran}
\IEEEoverridecommandlockouts

\usepackage{cite}
\usepackage{fancyhdr}
\usepackage{amsmath,amssymb,amsfonts}
\usepackage{algorithmic}
\usepackage{graphicx}
\usepackage{textcomp}
\usepackage{xcolor}
\def\BibTeX{{\rm B\kern-.05em{\sc i\kern-.025em b}\kern-.08em
    T\kern-.1667em\lower.7ex\hbox{E}\kern-.125emX}}

\usepackage{algorithmic}
\usepackage{graphicx}
\usepackage{textcomp}
\usepackage{xcolor}
\usepackage{soul}

\fancypagestyle{plain}{% copies "fancy" over "plain"
  \fancyfoot[C]{\thepage}% you can add edits that won't affect "fancy" but only "plain"
}

\usepackage{tabularx}
\usepackage{makecell}

\usepackage{pifont}

\def\BibTeX{{\rm B\kern-.05em{\sc i\kern-.025em b}\kern-.08em
    T\kern-.1667em\lower.7ex\hbox{E}\kern-.125emX}}

\author{
\IEEEauthorblockN{Philippos Papaphilippou}
\IEEEauthorblockA{
\textit{School of Electronics and Computer Science, University of Southampton, United Kingdom}\\
P.Philippos(a)soton.ac.uk}}

\usepackage{balance}  
\usepackage{easy-todo}
\usepackage[hidelinks]{hyperref}

\usepackage[shortcuts]{extdash}
\usepackage{setspace}

\begin{document}

\title{LUTstructions: Fast-Reconfigurable\\ FPGA-Based Instructions
}%
\maketitle

\begin{abstract}

 General-purpose processors %
 feature a limited number of instructions based on an instruction set. They can be numerous, such as with vector extensions that include hundreds or thousands of instructions, but this comes at a cost; they are often unable to express arbitrary tasks efficiently. This paper explores the concept of having reconfigurable instructions by incorporating reconfigurable areas in a softcore. It follows %
a relatively new %
computing paradigm
for seamlessly loading instruction implementation-carrying bitstreams from %
main memory. The resulting softcore is entirely evaluated on an FPGA, essentially having an FPGA-on-FPGA for the instruction implementations, with no notable operating frequency overhead. This is achieved with a custom FPGA architecture, %
 which is tailored towards low-latency for custom instructions %
and wide reconfiguration, as well as a soft implementation for the purposes of architectural exploration. All code is open-source to foster further research on reconfigurable instructions.
\end{abstract}


\section{Introduction}
\thispagestyle{plain}
With the rise of adoption of machine learning and artificial intelligence, incorporating accelerators is now considered integral to computers ranging from edge devices \cite{10829854} to supercomputers. A notable example can be seen by contrasting the historical percentages of the top 500 supercomputers that have accelerators. This number has increased from 29.2\% in June 2020 to 53\% in June 2025, with 19 out of the top 20 now all featuring accelerators \cite{top500}. In other words, even with advanced and wide vector extensions \cite{cebrian2020scalability,wilkinson2022initial}, general-purpose processors have failed to take on accelerators for raw compute.

Debatably, this change has happened too fast for the computing industry to adapt. %
Two different technologies; CPUs and accelerators are mostly developed separately, reflecting this division at the system-level \cite{10829854}. For instance, the highest-end GPUs and FPGAs are treated unfairly when it comes to acquiring main memory performance, since it comes through PCIe \cite{ren2025enabling}. This system heterogeneity complicates data-intensive FPGA designs %
\cite{martinelli2025bridging}, and creates different %
forms of redundancy in hardware, as with the requirement for  local memories for achieving processing with %
high bandwidth \cite{9460576}.

In order to address this heterogeneity challenge, a new computing paradigm is currently emerging. That is to bring accelerators and general-purpose computing as close together as possible. A prime example is how Apple silicon and Ryzen APUs are considered a highly cost-effective way to run large-language models today \cite{benazir2025benchmarking}. Due to their unified memory, hundreds of Gigabytes can more easily be allocated to the model. %
More related to this paper, there have been relatively recent attempts to bring FPGAs closer to the processor. %
Such paradigm shifts are expected to bloom, and this is reflected in the acquisition of leading FPGA companies by CPU companies such as Xilinx from AMD. Nonetheless, flagship
architectures like AMD Versal still involve accelerators in heterogeneous platforms \cite{mhatre2025performance} with shortcomings in complexity. %
Companies like QuickLogic as well as academic projects \cite{moser2024stitching} are now heavily invested in embedded FPGAs that can also implement instructions, but these are restricted for embedded use and %
infrequent configuration. %
Future processors with on-demand dynamic reconfiguration %
are predicted to be faster and more energy efficient than what is currently possible \cite{arc22fpgaext}.

The general \textit{\textbf{research question}} addressed in this work 
is whether  dynamic reconfiguration of FPGA-based instructions can happen fast enough so that they behave nearly indistinguishable from hardened instructions. In doing so, there are some critical challenges: 
\vspace{0.2em}
\paragraph*{\textbf{C1.}{ Reconfiguration latency}}
FPGAs are currently not optimised for fast reconfiguration \cite{versatile}. This becomes of crucial importance in scenarios with frequent context switching, or simply using multiple custom instructions. The additional latency to reconfigure the fabric happens whenever the instruction opcode alias differs from the currently programmed instruction, %
effectively becoming
an instruction \textit{implementation miss}. See section \ref{preconf} for numerical examples using current reconfiguration techniques. %
\paragraph*{\textbf{C2.}{ Instruction latency}} another challenge is the latency %
of the FPGA-based instructions %
for execution (i.e. \textit{implementation hit}). When compared to other functional units inside the core, this needs to remain in the same order of magnitude as hard instructions \cite{arc22fpgaext}. This can be challenging to achieve due to discrepancies in the %
operating frequency (challenge \textit{\textbf{C3}}) and any control-related logic that may delay the instruction execution. Ideally, %
custom instruction implementations would also %
be pipelinable to avoid blocking behaviour (backpressure) between consecutive calls and maximise the throughput.

\paragraph*{\textbf{C3.}{  Operating frequency}}
Due to the technological ``disadvantage'' of FPGAs when compared to hard logic, designs may exhibit up to an order of magnitude drop in the operating frequency. %
This performance bottleneck could be expected to migrate over reconfigurable instructions, when an existing FPGA architecture is used. %
The logic of an instruction could be too simple to outweigh this limitation through the flexibility and parallelism that comes with FPGAs. Similarly, different designs may yield different operating frequencies for custom instructions, and the clock domain crossing would either need to complicate the processor or be simplified by setting a ``safely'' low frequency, also impacting the throughput. %

\paragraph*{\textbf{C4.}{ Architectural exploration}} 
Modelling an architecture that combines modern hardened cores with dynamically-reconfigurable regions is challenging, %
because each of the two technologies has different physical and behavioural properties. On one hand, traditional computer architecture research has led to various time-saving techniques for estimating processor performance by reducing simulation time \cite{sabu2022looppoint}. On the other hand, FPGA research relies on FPGA-based prototypes that allow more complete executions for additional realism, such as the interaction with DRAM \cite{manev2019unexpected}. However, when it comes to combining the two, existing higher-level abstractions %
do not include fine-grain reconfiguration \cite{karandikar2019using}, and lower-level models such as through fabrication may need to be prohibitively large and
 expensive to capture %
 system-level behaviours \cite{arc22fpgaext}. %

\begin{figure}[t!]
\centering
\includegraphics[width=0.45\textwidth, trim=0 5 0 0]{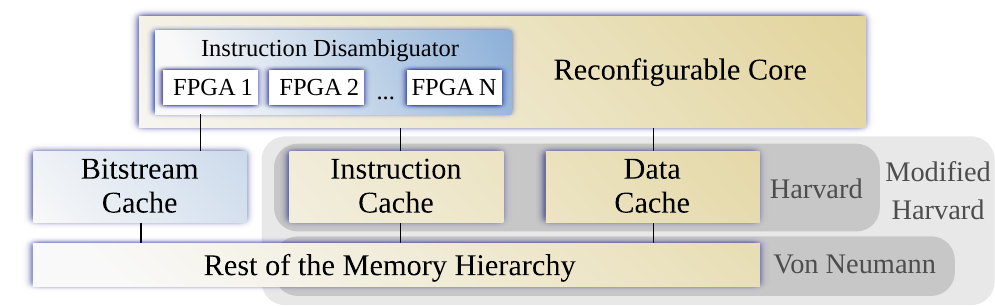}
\caption{FPGA-extended  modified-Harvard
computer architecture. %
}\label{fpgaext}
\vspace{-1.em}
\end{figure}

\vspace{0.2em}
\paragraph*{\textbf{Motivation}} This work is inspired by the FPGA-extended computer architecture \cite{arc22fpgaext}, %
which is summarised in figure \ref{fpgaext}. %
This architecture aims at achieving FPGA performance in general-purpose processors. It combines small FPGAs working as instructions (managed by the instruction disambiguator) inside each CPU core that are programmed dynamically with bitstreams residing in the same memory space. A specialised cache (bitstream cache) resides alongside the instruction and data caches to provide the bitstreams with a high bandwidth on demand. %
It is shown that a low-enough reconfiguration latency can behave similarly to having all utilised instructions hardened, even with multiprocessing in modern operating systems. %
The experiments %
 introduce an artificial latency on every opcode miss, to cover different potential scenarios for future implementations. %
While the concept is uniquely promising,
it does %
not elaborate with a design for the reconfigurable regions. %
As demonstrated in the literature review (section \ref{rw}), there is a growing gap for a modern and open  solution that combines such architecture homogeneity principles. %

\vspace{0.2em}
The presented research focuses on %
facilitating dynamically-loading reconfigurable instructions through %
a specialised FPGA architecture that addresses the aforementioned challenges.
The list of research contributions is as follows:
\begin{itemize}
\item %
LUTstructions, a custom FPGA architecture optimised for fast reconfiguration targeting dynamically-loading soft instructions.
\item The fastest FPGA-on-FPGA architecture to date, with soft designs having similar performance attributes to conventional FPGA pipelined designs.
\item Open-source implementation and end-to-end software support at \textit{\url{https://philippos.info/lutstructions}}. %
\item The first end-to-end implementation of the FPGA-extended modified-Harvard computer architecture. %
\end{itemize}

\begin{table*}[h!] 
\small
\vspace{-0.5em}
\footnotesize	
\caption{Qualitative comparison with representative %
research on reconfigurable instructions in chronological order.} 
\label{tab0}
\centering
\setstretch{-0.0}
\setlength{\tabcolsep}{2pt}
\begin{tabular} {l |c c c |c c c | c c |c c c }
&\multicolumn{3}{c|}{Reconfigurability}&\multicolumn{3}{c|}{Integration}&\multicolumn{2}{c|}{Latency (cycles)}&\multicolumn{3}{c}{Prototype}\\%
\multicolumn{1}{c|}{Work}&\thead{Run-time\\[0.2em] reconfigurable}&\thead{Logic\\ granularity}&\thead{Avoids\\[0.2em] control \\[0.2em]takeover}&\thead{Same\\[0.2em]ISA}&\thead{Same\\[0.2em]address\\[0.25em] space}&Cached&Configuration&Execution&ISA&Platform&\thead{Open\\[0.4em]source}\\%
\hline
PRISC \cite{razdan1994prisc}&No&FPGA&\ding{51}&\ding{51}%
&\ding{51}&N/A&100-561&$\ge4$&MIPS&\thead{Performance\\[0.2em]model}&No\\[-0.2em]%
DISC \cite{wirthlin1995disc}&No&\thead{(external\\streams)}&No&\ding{51}&\ding{51}&N/A&N/A&\thead{$\sim$5-10\\[0.2em] (scheduling)}&toy&Simulation&No\\[-0.2em]%
RaPiD \cite{ebeling1996rapid}&No&\thead{FPGA\\[0.2em](coarse-grain)}&No&N/A&No&N/A&\textit{\thead{not\\known}}&\textit{\thead{not\\known}}&N/A&Simulation&No\\[-0.2em]%
Garp \cite{hauser1997garp}&\ding{51}&\thead{Instruction\\[0.2em]combinations}&No&No&\ding{51}&No&\textit{\thead{not\\known}}&\textit{\thead{not\\known}}&\thead{MIPS \&\\[0.2em] custom}&Simulation&No\\[-0.2em]%
MorphoSys \cite{lu1999morphosys}&\ding{51}&CGRA&\thead{No\\[0.2em](DMA)}&No&No&(storage)&8&37&\thead{MIPS-like \&\\[0.2em] custom (DMA)}&ASIC%
&No\\[-0.2em]%
Chimaera \cite{ye2000chimaera}&\ding{51}&FPGA rows&\ding{51}&No&\ding{51}&\ding{51}&\textit{\thead{not\\known}}&1-32&MIPS&\thead{Sim. (high-level),\\[0.2em] ASIC (fabric)}&No\\%
PipeRench \cite{goldstein2000piperench}&\ding{51}&\thead{Instruction\\[0.2em]combinations}&\ding{51}&\textit{\thead{not\\known}}&\textit{\thead{not\\known}}&N/A&1&\textit{\thead{not\\known}}&N/A&ASIC&No\\[-0.2em]%
ADRES \cite{mei2003adres}&No&CGRA&\thead{No\\[0.2em](VLIW)}&\ding{51}&No&N/A&\textit{\thead{not\\known}}&\textit{\thead{not\\known}}&toy (IMPACT)&Co-simulation&No\\[-0.2em]%
MOLEN \cite{vassiliadis2004molen}&\thead{No (one-time\\[0.2em] extension)}&CGRA&No&No&No&(storage)&\textit{\thead{not\\known}}&17-25&\thead{PowerPC \&\\[0.2em] custom}&FPGA&No\\[-0.2em]%
RISPP \cite{bauer2007rispp}&\ding{51}&FPGA&No&No&No&No&\thead{$\sim$90,000\\[0.2em]@100MHz}&5-24&toy (DLX)&FPGA&No\\[-0.2em]
Ordaz et al. \cite{ordaz2018soft}&\ding{51}&FPGA&\ding{51}&\ding{51}&No&No&\thead{$\sim$29,500\\[0.2em]@100MHz}&$\sim$1-5&RISC-V&\thead{Sim. (evaluation),\\[0.2em] FPGA (impl.)}&No\\
\emph{\textbf{LUTstructions}}&\ding{51}&FPGA&\ding{51}&\ding{51}&\ding{51}&\ding{51}&32&5&RISC-V&\thead{FPGA,\\[0.2em] ASIC (fabric)}&\ding{51}\\[-0.2em]

\end{tabular}
\setstretch{1}
\vspace{-1.5em}
\end{table*}

\section{%
Related work}\label{rw}

This section serves as a literature review on categorised existing and emerging technologies relating to the presented work on reconfigurable instructions, and how they compare as alternatives, inspiration or partial replacements. %

\subsection{Reconfigurable instructions}%

While historically there have been plenty of works introducing reconfigurable regions into %
or near %
CPU cores, %
they exhibit considerable limitations relating to dynamic reconfiguration for custom instructions. %
The absence of a unified computer architecture has debatably limited their adoption to only academic proof-of-concepts on embedded systems \cite{%
ahmed2011efpgas}. 

Table \ref{tab0} introduces a comparison of representative works on key areas that are deemed important for addressing the research challenges. %
Starting from the left, the approach needs to be \textit{run-time reconfigurable} which is the scope of the paper. The \textit{logic granularity} impacts the flexibility of the reconfigurable areas to express logic beyond traditional instructions. For a tighter integration, custom instructions can avoid \textit{taking over} the control of the CPU in the style of workload offloading. The modification on the base \textit{ISA} needs to be minimal and homogeneous, such as by accessing the same register file. When the bitstreams reside in the \textit{same address space} and are \textit{cached}, this enables %
the hardware to automatically facilitate the reconfiguration efficiently. The \textit{configuration} and \textit{execution} latencies correspond to challenges \textbf{\emph{C1}} and \textbf{\emph{C2}} respectively. The last three columns relate to challenge \textbf{\emph{C4}} and provide the \emph{ISA} of the prototype, its evaluation \emph{platform} as an indicator of readiness, and whether it is open-source. Challenge \textbf{\emph{C3}} is not directly discussed in this table, as the exact frequency also depends more on the technology of the time, but influences design choices such as the \emph{logic granularity}. This comparison is qualitative and indicative, as each work provides different levels of detail and these notions may be used differently. For instance, the configuration information may exclusively be expressed in the form of instructions that already pass through the instruction cache, so this would involve \emph{caching}. %

DISC \cite{wirthlin1995disc} is an early attempt to orchestrate accelerators within the CPU core's control flow, though it concerns external devices, and hence is orthogonal to instruction reconfigurability. PRISC \cite{razdan1994prisc} also introduces FPGA-levels of flexibility with a LUT-mesh, though its high reconfiguration latency is prohibitive, and it was only demonstrated with a theoretical performance model.  Garp \cite{hauser1997garp} introduces a systolic array-like structure to accelerate parts of existing code, but a rather high number of additional special instructions are required to manually manage the array. %
To overcome the reconfiguration challenges, additional works such as RaPiD focused on coarsening the fabric's architecture \cite{ebeling1996rapid}, but they have not been demonstrated as part of a reconfigurable core. 

Similarly, PipeRench \cite{goldstein2000piperench} further restricted the reconfiguration flexibility to only implement sets of existing instructions, in order to achieve fast reconfiguration. One advantage of PipeRench and similar works is that they come with a C compiler that automatically alters existing code to benefit from the reconfigurable region. These have arguably been abandoned in favour of instruction-level parallelism (ILP) in modern processors, since they have the common goal of optimising the utilisation of existing functional units. For more specialised applications, HLS could be seen as a modern alternative, though this often involves no instructions. %

Coarse-grained reconfigurable arrays (CGRAs) have also been utilised to implement reconfigurable instructions. For example, MOLEN \cite{vassiliadis2004molen} provides one of the most complete prototypes from the prior work by using an FPGA, %
as opposed to basing the evaluation on simulations. %
However, MOLEN requires the ISA to be manually extended with calls to kernels of interest for specialised applications before execution, so there is limited scope for run-time reconfiguration. ADRES \cite{mei2003adres} provides a more tightly-coupled integration with the base ISA, such as by accessing a common register file, as with the proposed solution. Despite this homogeneity, the reliance on a very long instruction word (VLIW) architecture enforces restrictions in the processor control flow, and VLIW is not considered general-purpose today. The principles behind ADRES are first introduced in MorphoSys \cite{lu1999morphosys}, which is more generalisable, such as by supporting run-time reconfiguration as context switching. As opposed to the presented approach, its ISA is less homogeneous as it requires a series of instructions relating to the encapsulated direct-memory access (DMA) engine to manually engage the CGRA. MOLEN and MorphoSys have local storage to store the configurations, which can have the effects of caching, though this is antithetical to the idea of unifying the address space and memory.

Chimaera \cite{ye2000chimaera} attempts to provide fast-reprogrammable FPGA-based instructions using a bitstream cache as well. Its fabric is mainly intended to be utilised by a compiler to map existing routines, and its granularity is finer than an FPGA, as it is divided into autonomous rows of limited functionality that can be combined. As an architecture it is fairly heterogeneous, such as by using a shadow register and relying on advanced compiler routines, whose complexity may be the reason behind the high-level simulation-based evaluation. RISPP \cite{bauer2007rispp} is another take on FPGA-based instructions, but by using the configuration facility of commercial FPGAs the programming time is too high for practical use for run-time reconfiguration, and this is at 0.9 milliseconds. Ordaz et al. \cite{ordaz2018soft} modernised this idea with a RISC-V softcore, but due to the reliance on ICAP through partial reconfiguration the run-time reconfiguration latency is at 0.295 milliseconds. The authors' suggested workaround is to use prefetching of the bitstreams. However, with an overhead of 29.5K cycles at 100 MHz, %
the claimed run-time reconfiguration would only suffice for extreme instruction usage patterns, where this latency would meaningfully be spent by other instructions or tasks. %

As can be observed from table \ref{tab0}, none of the competing solutions have the required combination of features to be able to address %
 the identified research challenges. There is the prominent trade-off between the granularity of the fabric and the reconfiguration capabilities (latencies, run-time reconfiguration and automation). There are works that come close to having most desired features supported such as PRISC \cite{razdan1994prisc} and Chimaera \cite{ye2000chimaera}, but these are evaluated using a performance model and a high-level simulation respectively, highlighting the importance of this research area, such as by being relatively conceptual. In contrast, the proposed solution is fully-implemented in Verilog and purposely designed to address the challenges in a natural and homogeneous way by following a general computer architecture \cite{arc22fpgaext}. It is also the only entry in the table that is open-source, with an aim to foster ongoing and future research on reconfigurable instructions.

\subsection{Embedded FPGAs (eFPGAs)}\label{efpga} 

An emerging technology is eFPGAs and is already being used to implement custom instructions \cite{koch2021fabulous}. Being more contemporary, eFPGA-based approaches are more frequently open-source, as with FlexBex \cite{flexbex} and Greyhound \cite{moser2025greyhound} that also harness the openness of RISC-V. %
Still, they show noticeable limitations for implementing dynamically-reconfigurable instructions. These are inline with prior works utilising larger designs on a discrete FPGA fabric that takes over the control from the host processor, such as RISPP \cite{bauer2007rispp} and PRISC \cite{razdan1994prisc}. %

In general, eFPGA-based research currently focuses on modular integration and concerns larger areas and slow reconfiguration \cite{abideen2025efpga}. %
This obstacle is expected, since they inherit the fundamental limitation of using a standard %
 configuration protocol, which handicaps their dynamic reconfiguration capabilities. The decoupled aspect of eFPGAs shifts away from the dynamically-reconfigurable instruction paradigm due to the high reconfiguration latency and system heterogeneity.
Regardless of the common %
integration conventions, %
a LUTstruction %
could still be considered an eFPGA in loose terms.

\subsection{Partial reconfiguration}\label{preconf} 
Partial reconfiguration \cite{vipin2018fpga} is supported by major FPGA vendors in their latest architectures to allow segmenting the FPGA fabric into multiple reconfigurable ones, so that modular designs can be reprogrammed during runtime \cite{vaishnav2020fos}.

The proposed solution could be demonstrated to some extent with the use of partial reconfiguration on a similar %
platform. %
However, this would not allow for the exploration of the custom %
FPGA architecture explored here (challenge \emph{\textbf{C4}}), as the logic elements would be hardened and become equivalent to those used for the softcore. Importantly, modern partial reconfiguration inherits limitations of proprietary platforms such as the %
configuration rate and less area-flexibility. %

For instance, the Internal Configuration Access Port (ICAP) used to facilitate partial reconfiguration on 
AMD devices %
is capped at 100 MHz and is 32-bit wide \cite{icap}. %
As a back of the envelope calculation, if an instruction bitstream is 8 KiB, loading it through ICAP at 100 MHz would take 2048 FPGA cycles (challenge \emph{\textbf{C1}}). To keep things in perspective, %
one of the most costly SIMD intrinsics is for 8 packed 64-bit floating point division in AVX-512, and takes up to 23 CPU cycles to execute (\texttt{\_mm512\_div\_pd} \cite{intel}).

The state-of-the-art in fast reconfiguration builds on partial reconfiguration, such as by overclocking ICAP \cite{versatile}. Still, this %
would be impractically slow at 1.4 GB/s over the achieved 38.4 in our P=16 configurations (see sections \ref{parrec} and \ref{dses}) that avoid standardised reconfiguration to solve challenge \emph{\textbf{C1}}. %

\subsection{Specialised FPGAs and overlays}\label{overl}

The presented evaluation involves an FPGA-on-FPGA. Zuma is a representative framework for inferring virtual FPGAs on FPGAs \cite{brant2012zuma}. It does not focus on performance%
, and 
 does not mention any operating frequency. 
Subsequent adaptations have reported sub-1 MHz for a Zuma region \cite{wiersema2014embedding} (challenge \emph{\textbf{C4}}).
In contrast, our presented design for S=1 achieves 1058.15 MHz on Alveo V80. %

With respect to the %
data movement, the presented FPGA architecture is somewhat reminiscent of systolic arrays for more specialised applications \cite{xu2023survey}. Involving logic elements for routing inside the fabric %
(see design choice 2 of section \ref{desfa}) is partly inspired by \emph{minimal FPGAs} \cite{rotfpga} and could be classified as such. Later works on %
overlays have strongly shifted the focus to domain-specific applications \cite{liu2022overgen} whose reconfiguration is less flexible and adaptable for our use case. 

There are also special-purpose FPGA architectures such as Triptych \cite{hauck1992triptych} and RaPiD \cite{ebeling1996rapid} that target %
modularised tasks including reconfigurable instructions. With the continuous improvement of recent attempts for an open-source FPGA such as with OpenFPGA \cite{tang2020openfpga}, more unconventional architectures are expected to bloom \cite{youssef2025lazagna}. Such research is not %
orthogonal to the presented FPGA, as future work could investigate emerging concepts for adoption in the topology of the fabric. %

\section{LUTstructions design}%
\label{desfa} %

A novel field-programmable gate-array (FPGA) 
architecture is presented that specialises for custom instruction implementations, such as by being small and fast. In its present state, it is also optimised for integration within softcores, to %
be able to efficiently implement an ``FPGA-on-FPGA''. %

The fabric is designed with dataflow computing principles in mind. %
 The unique design choices to specialise the fabric for instructions are as follows:

\begin{enumerate}
\item \emph{No backward movement:} the information figuratively only propagates from left to right. %
 The instruction inputs are on the left (e.g. two \(W\)-bit values), and the output value is on the right. In this way, the design can be pipelined, and all computation is assigned a fixed latency, the length of the fabric \(Y\) (\textit{Y\(=\)W\(=\)}32 in the %
 methodology). %
 This mainly addresses challenge \textbf{\emph{C2}}.
 \item \emph{Look-up tables responsible for routing:} moderately complex logic can be expressed without relying on dedicated routing logic, due to the diagonal outputs that can be used to propagate the signals. This %
simplifies FPGA implementation for architectural exploration to address challenges \textbf{\emph{C3}} and \textbf{\emph{C4}}, since multiplexers are generally considered costly as soft logic. %
Specifically, it uses LUT4\_4 look-up tables, i.e.\ with 4 inputs and 4 outputs, as shown in figure \ref{lut4_4}. The naming is inspired by AMD's LUT6\_2 primitives. One LUT4\_4 is equivalent to four LUT4 sharing the same inputs. 
\begin{figure}[h!]
\centering
\includegraphics[width=0.25\textwidth, trim=0 7 0 0]{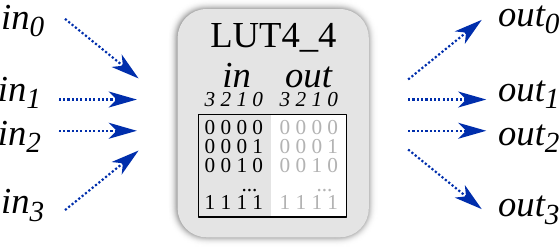}
\caption{LUT4\_4: Look-up table with 4 inputs and 4 outputs. %
}\label{lut4_4}
\vspace{-1em}
\end{figure}

\item %
\emph{No registers:} the modelled logic cannot use registers, and all state shall use the core's traditional registers, addressing challenges \textbf{\emph{C2}} and \textbf{\emph{C4}}. This is also to adhere to conventional programming models and make it instruction-specific, though future research includes experimentation with stateful instructions (instructions that can hold states between their calls). This assumption also minimises potential safety and security concerns in more advanced micro-architectures by effectively becoming a functional unit of a fixed pipeline length.

\end{enumerate}

\begin{figure}[h!]
\centering
\includegraphics[width=0.45\textwidth, trim=0 10 0 10]{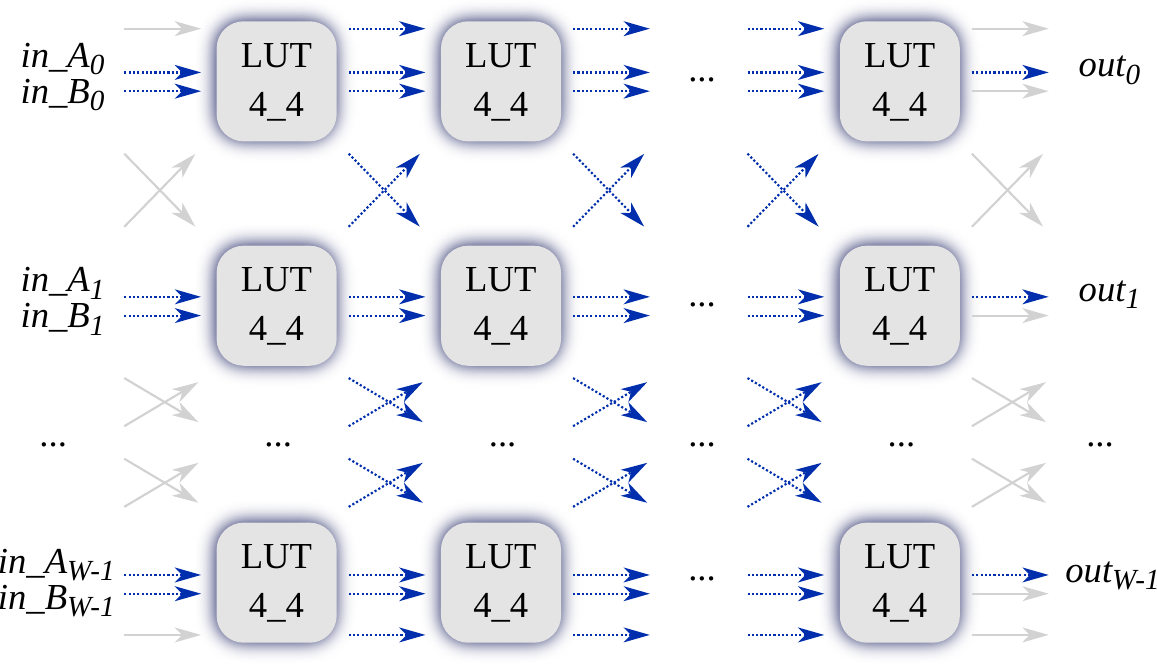}
\caption{FPGA architecture %
for reconfigurable instructions.%
}\label{fabric}
\end{figure}

The resulting architecture  is summarised in figure \ref{fabric}. A few additional optimisations are needed to achieve low pipeline and reconfiguration latencies for implementing instructions.

\subsection{Register placement}\label{regpla}

The logic blocks of conventional FPGA architectures include a register component, which is optionally enabled per logic element to be able to express circuits that hold small states internally. The enablement of those registers happens at the configuration stage, and this directly impacts the critical path and %
operating frequency. Since the proposed architecture currently includes no registers in its logic blocks, and the data only move in one direction, the critical path would be directly proportional to the depth (\textit{Y}) of the instruction fabric. This can become a limiting factor, when \textit{Y} is sufficiently large, and for our exploration with a square-shaped fabric (\textit{W=Y}) this is the case. The results of section \ref{dses} elaborate on the timing effects of having no registers, for \textit{S=32}, where \(S\) also corresponds to the critical path length.

\begin{figure}[h!]
\centering
\includegraphics[width=0.46\textwidth, trim=0 20 0 15]{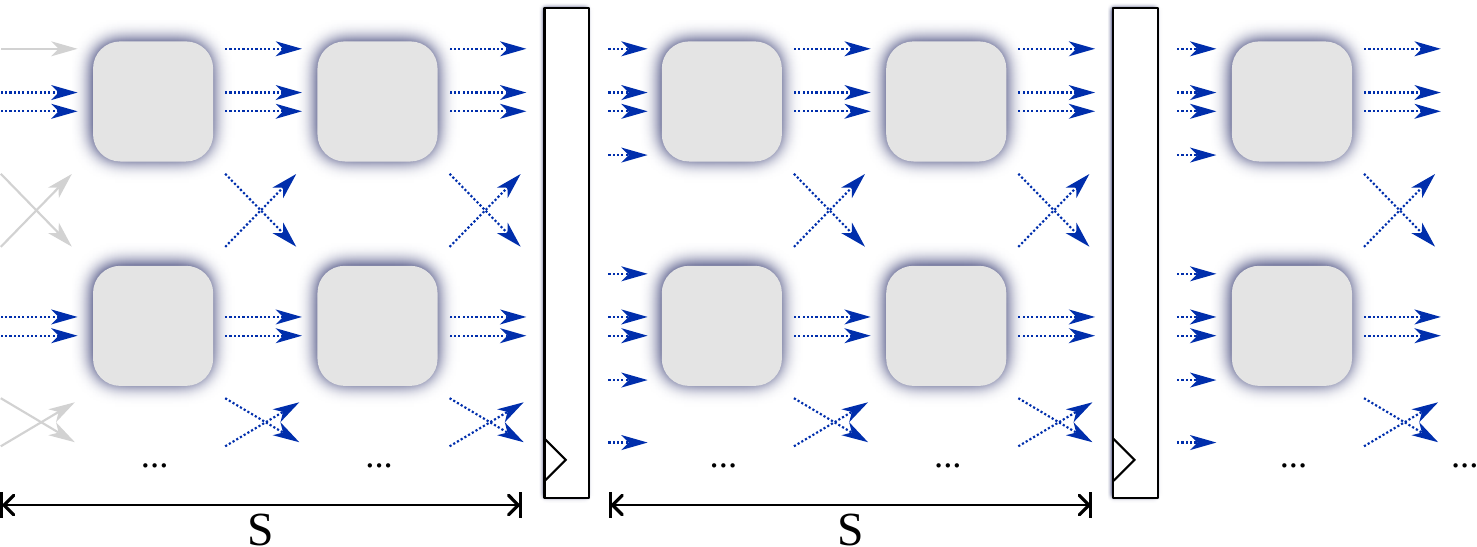}
\caption{Placing compulsory registers on every S LUTs, %
S=2.
}\label{fabricf}
\end{figure}

Figure \ref{fabricf} introduces the register placement optimisation, and directly targets challenge \textbf{\emph{C2}}. There is a compulsory register across a LUT column on only every \(S\)th column. This approach follows a %
dataflow %
approach, and is analogous to register retiming as found in commercial FPGA toolchains. %
However, this %
is applied to the soft fabric as a design to be implemented on the real FPGA. This is done regardless of the intended logic that would be expressed as custom instructions later, and ensures passing universal timing constraints for all possible bitstreams-instructions targeting the soft FPGA. %

This approach also emphasises on %
achieving a high-performing FPGA-on-FPGA setup (addressing \textbf{\emph{C4}}), and could be superfluous for simpler instructions. %
On the other hand, it simplifies the placement of the logic by completely disregarding timing analyses for custom instructions %
after the fabric has been mapped to real hardware. The fine-grain control of the critical path using the \(S\) parameter allows effortless %
increase of the operating frequency (resolving challenge \textbf{\emph{C3}}) for arbitrarily complex circuits expressible by a LUTstruction. %

\subsection{Configuration parallelism}\label{parrec} %
The %
fabric is programmed in a pipelined fashion, and the wires used for logic are reused for propagating the configuration from column to column. When the %
fabric is reset to receive a new configuration, each LUT4\_4 is operating in bypass mode %
(i.e.\ \textit{out$_0$}$\leftarrow$\textit{in$_0$} etc.),  as indicated in the example configuration of figure \ref{lut4_4}. Then, the bitstream information is propagated until it reaches the LUTs of the row that is being programmed, starting from the rightmost column.

Notice how the diagonals are alternating in a zigzag pattern when operating in bypass mode. This is intentional to avoid shifting the configuration bits too much. Effectively, starting from the left where the bitstream is loaded, every second column has its diagonal wires swapped with those of the neighbours. This is easily resolved at the bitstream generation stage, where every 4 bits belonging to a LUT of an odd column have their first and last bits swapped with the adjacent wires of their neighbours. This excludes some edge cases like the first bit of the first LUT and last bit of the last LUT. %

\begin{figure}[h!]
\centering
\includegraphics[width=0.45\textwidth, trim=0 8 0 0]{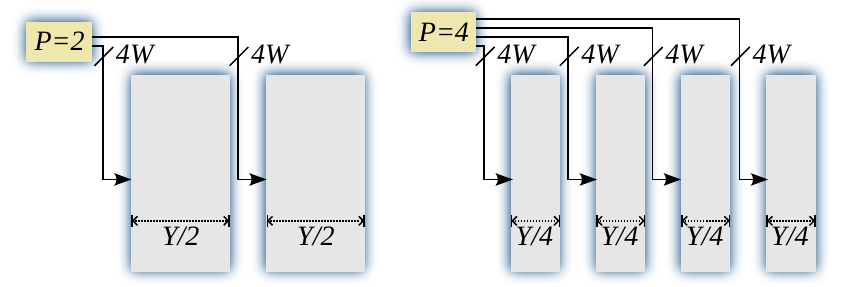}
\caption{Parallel configuration of same-sized fabric chunks.%
}\label{parallelconf}
\vspace{-0.5em}
\end{figure}

In order to further increase the configuration width, the %
fabric can be divided into P equally-sized segments, where P is referred to as configuration parallelism. The range of supported values for P is all powers of two up to half the fabric depth (Y/2) due to the bit swapping workaround for odd columns. Though further specialisation can be trivial, such as to discard the aforementioned workaround, if all columns are programmed in parallel. %
This optimisation is visualised in figure \ref{parallelconf}, and resolves challenge \textbf{\emph{C1}}.

\subsection{Bitstream cache}
In order to support the wide reconfiguration without affecting the datapath requirements for the remainder of the architecture, %
a custom cache is required (associated with challenge \textbf{\emph{C1}}). This cache is able to provide \(4W\times P=128\times P\) bits per cycle of the corresponding bitstream to the instruction disambiguator, whenever there is an instruction implementation miss. Refer to section \ref{simod} for more insights on the pursued implementation, while noting that the core's existing cache %
conventions %
can be influential on the final design choices. %

\section{Methodology}

Since the project involves an alternative computer architecture, multiple software and hardware aspects of the system are concerned, in order to showcase an end-to-end working solution%
. These include the core architecture and micro-architecture, programmability, instruction bitstream generation, and implementation on real hardware using an FPGA. %

\subsection{RISC-V adoption and GNU toolchain}\label{riscv}

The implementation of the proposed solution is consistent with the RISC-V specification, and does not require the definition of a new instruction type. This is because RISC-V promotes custom instruction development by providing four pre-defined opcodes for custom instructions, namely \textit{custom-0} to \textit{custom-3}. These opcodes are highly-flexible with respect to aliasing. The unused parts of the instruction can be used to multiplex a relatively high number of custom instructions using a single base %
opcode. The reconfigurable instructions are currently evaluated as functional units in the sense that they adhere to the conventions of a typical use of an arithmetic logic unit (ALU). Therefore, no special registers or control logic is required from RISC-V's point of view.

\begin{figure}[h!]
\centering
\includegraphics[width=0.48\textwidth, trim=0 7 0 5]{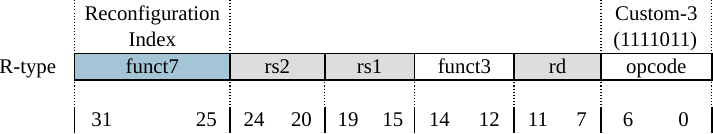}
\caption{%
Adopting the R-type instruction format.
}\label{itypes}
\end{figure}

As shown in figure \ref{itypes}, LUTstructions %
are here evaluated as R-type instructions, i.e.\ accepting two register values and returning one. The \textit{funct7} field originally used for aliasing is reused to disambiguate between different instruction configurations, totalling \(2^7=128\) bitstreams. The current assumption is that the bitstreams reside continuously in main memory. %
Hence, \textit{funct7} acts as the reconfiguration index and denotes a memory location, in the same way a program counter can seek instructions originating from a binary. %

Similarly, the \textit{funct3} field can be reused as an input to the logic to alter the bitstream behaviour. This is useful when combining up to  \(2^3=8\)  instructions in a single bitstream (totalling 1024 custom instructions per base opcode), and this is analogous to how similar instructions in RISC-V extensions tend to reuse the same circuitry. According to the requirements, this scheme could be further adapted such as to include more bitstream locations by reserving more 
opcodes. %

What requires a minor modification is the current version of RISC-V GNU Compiler Toolchain. This is for GCC to be able to associate custom instruction names with the corresponding opcode, alias and instruction type, to pass the correct arguments. The corresponding changes are added as instruction \textit{name}, \textit{match}, and \textit{mask} fields for every new supported instruction. Instead, generic names can also be given as placeholders, such as \texttt{c0001} for the bitstream 1, to avoid recompilation on every addition of a new custom instruction. 

\begin{figure}[h!]
\centering
\includegraphics[width=0.48\textwidth, trim=0 8 0 5]{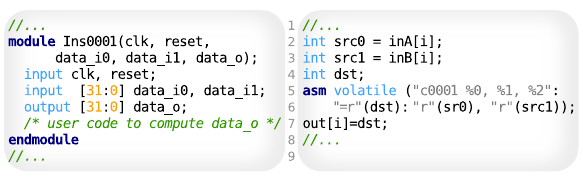}
\caption{%
Programmability example: Verilog template (left), C code (right).
}\label{progr}
\end{figure}

This arrangement assumes the software programmers will use inline assembly in a similar fashion to Intel's SIMD intrinsics  \cite{intel}, though future work could focus on automating the inference of those instructions through generic C/C++ code. Figure \ref{progr} demonstrates how the corresponding Verilog instruction description in the provided template can be ``called'' from a simple C program with the help of inline assembly.

Note that the LUTstruction
fabric can be used beyond RISC-V and specific instruction types, such as with up to 4 input and 4 output 32-bit registers for LUT4\_4 (\(W=32\)), as well as for SIMD instructions (\(W=\text{VLEN}\), the vector register size).

\subsection{RISC-V softcore}\label{simod}

The RISC-V design behind the presented methodology is %
\cite{simodense0}. Its base version implements the RV32IM specification, i.e.\ the base 32-bit integer ISA plus the multiplication extension. %
The reasoning behind this selection is that %
being optimised for SIMD instructions, it has a focus on wide data paths and larger caches for implementation on FPGAs. Having an FPGA-focused cache hierarchy with wide data paths is useful %
for being able to support the proposed bitstream cache for achieving wide and fast reconfiguration. %

\begin{figure}[h!]
\centering
\includegraphics[width=0.43\textwidth, trim=0 5 0 0]{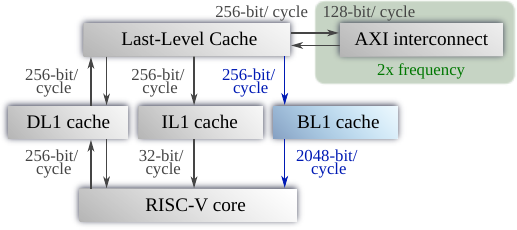}
\caption{Memory organisation %
 involving a bitstream cache.}\label{hierarchy}
\vspace{-0.7em}
\end{figure}

As illustrated in figure \ref{hierarchy}, the core is extended with the bitstream cache (BL1) at level 1 alongside the data (DL1) and instruction (IL1) caches. At level 2, there is a shared cache (LLC) that is connected to all level 1 caches. The tested implementation has 256-bit/cycle datapaths between most entities and to main memory%
, with the exception of the link between BL1 and the core which is wider, such as 2048-bit. The latter design choice is explored in the presented exploration of section \ref{dse}, as it directly relates to the effectiveness of the proposed solution, though the overall core remains highly parameterisable with regard to the bus configurations. 

The bitstream cache (BL1) is implemented in block RAM, since its blocks are relatively large %
to be able to fit the 8 KiB bitstreams of the study. A strobe mechanism is implemented to enable accessing those bitstreams in chunks to enable an efficient use of the primitive BMEM blocks. %
The width of these chunks is the maximum between the cache width of the upper level %
and the configuration width, %
to fully-support the bandwidth of both fetching and supplying the bitstreams. %

\subsection {Indicative system setup}%

\begin{figure}[h!]
\centering
\includegraphics[width=0.37\textwidth, trim=0 8 0 5]{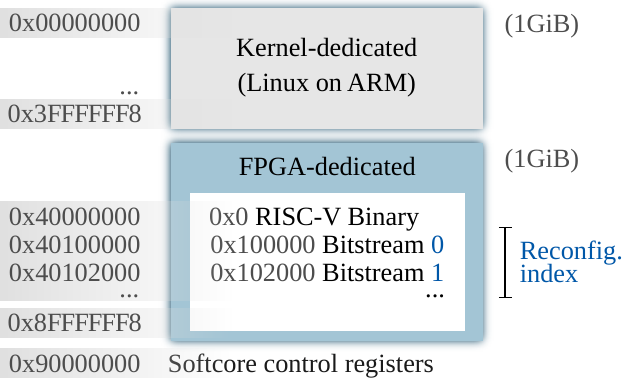}
\caption{Physical memory address space in validation setup. %
}\label{physicalmem}
\end{figure}

The design is also validated separately on the Ultra96 FPGA board featuring the ZU3EG device. %
The SoC combines 4 ARM cores running Linux and an FPGA. Its 2 GiB of memory is here divided in two segments. %
The 1st GiB is dedicated to Linux, and the 2nd GiB is entirely allocated to the RISC-V softcore. The softcore's memory address space is \textit{OR}ed to \texttt{0x40000000} right before communicating through AXI, to be mapped on to this 2nd GiB of the main memory. %
This mapping is summarised in figure \ref{physicalmem}.

\begin{figure*}[h!]
\centering
\includegraphics[width=0.99\textwidth, trim=0 8 0 0]{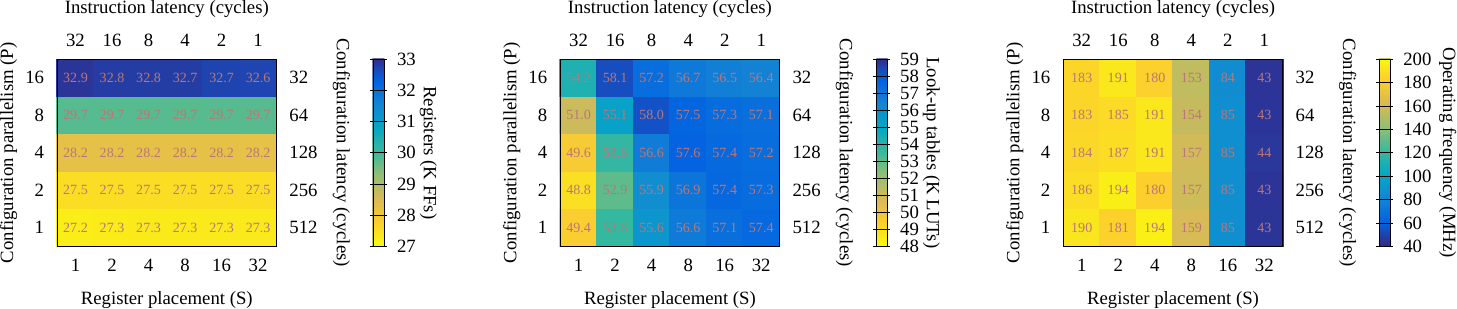}
\caption{Minimising reconfigurable instruction latencies %
within a RISC-V softcore, and their impact on FFs (left), LUTs (middle), and $f_{max}$ (right).}\label{hmap}
\vspace{-0.7em}
\end{figure*}

Before execution, a C program in Linux on ARM is responsible for loading the RISC-V binary and instruction bitstreams to dedicated locations in the FPGA memory space. As with the start address denoting where the RISC-V program starts, the user  provides the bitstream library address within the RISC-V softcore's memory. The reconfiguration index is used to map a high number of bitstreams in consecutive memory starting from the example bitstream library address of \texttt{0x100000}.  %
This is done in order for the softcore's memory system to automatically be able to fetch the instruction implementations on-demand, %
as it already does for data and instructions.

\subsection{Synthesis and routing}

The reconfigurable instruction functionality is expressed in Verilog within an encapsulating Verilog module (see figure \ref{progr}, left). The Verilog module has the input data at its input, and expects the resulting value at its output, similar to how an ALU would be interfaced. %

The first step is to synthesise the logic into \textit{LUT4\_4}s, i.e.\ LUTs with 4 inputs and 4 outputs. Most synthesis tools including \textit{abc} do not support gates or look-up tables with more than 2 outputs, and return the corresponding error. This is unless a more substantial coding effort is attempted. In order to achieve the synthesis, a custom FPGA architecture description is written in an \textit{xml} file for use with Verilog-to-Routing, %
which still uses \textit{abc} internally. The \textit{LUT4\_4} is modelled similar to fracturable LUTs and is equivalent to 4 \textit{LUT4}s having their inputs shared and their outputs unified. %

The final routing is done by a custom router that uses a %
search algorithm. %
It accepts the logic in the Berkeley logic interchange format (BLIF) from the latter step. It is an iterative approach where the \textit{LUT4\_4}s are mapped, then the remaining \textit{LUT4\_4}s are repurposed for routing purposes, as well as to express \textit{LUT3}s etc. It finally generates the instruction bitstream, also taking into account the bit swapping and the P-value (section \ref{parrec}), appropriately interleaving the corresponding bit sections for %
parallel configuration. This fully-open source approach is practically platform-independent, enabling potential reuse by the softcore to reconfigure itself. %

\section{Evaluation %
}\label{dse}

At face value, the presented evaluation is mostly a design space exploration (DSE) %
that can be followed towards the final design stages of chip design. %
The main aim is to demonstrate the effectiveness of %
LUTstructions %
and the presented optimisations including the FPGA-on-FPGA aspect. At the same time, it implies the feasibility of the model computer architecture \cite{arc22fpgaext} by materialising a complete prototype on an FPGA.

Due to the high number of variables that exist with essentially %
optimising a %
core with added reconfigurability, the DSE is only exhaustive with respect to sets of variables of interest. Thus, a baseline configuration is selected, and is summarised in table \ref{tab1}. Starting from the left, the instruction disambiguator (ID) has 2 LUTstruction slots (fabrics) that store a 8KiB-sized bitstream each. These are fed by the bitstream cache (BL1), which carries 16 bitstreams, one for each of its %
blocks. %

\begin{table}[h!] 
\vspace{-0.4em}

\small
\footnotesize	
\caption{Softcore baseline configuration.} 
\label{tab1}
\centering
\setlength{\tabcolsep}{2.5pt}
\setstretch{0.5}
\begin{tabular} {c|c c |c c | c c c | c c c  }
ID&\multicolumn{2}{c|}{BL1}&\multicolumn{2}{c|}{IL1}&\multicolumn{3}{c|}{DL1}&\multicolumn{3}{c}{LLC}\\
slots&sets&block&sets&block&sets&ways&block&sets&ways&block\\%
&&(bits)&&(bits)&&&(bits)&&&(bits)\\
\hline
&&&&&&&&&&\\
2&16&65536&64&256&16&4&256&16&4&16384\\[0.5em]%
(=16KiB)&\multicolumn{2}{c|}{(=128KiB)}&\multicolumn{2}{c|}{(=2KiB)}&\multicolumn{3}{c|}{(=2KiB)}&\multicolumn{3}{c}{(=128KiB)}\\
\end{tabular}
\setstretch{1}
\end{table}

The main evaluation platform for this section is AMD Alveo V80 to enable a wide design exploration due to the ample available resources %
(2.6M %
LUTs and 5.1M FF). Otherwise, for resource constrained devices Vivado would yield more variation, as it triggers more heuristics %
under pressure.

\subsection{Minimising %
instruction latencies}\label{dses}

The main exploration investigates the impact of the optimisations for \textit{parallel reconfiguration} (section \ref{parrec}) and \textit{register placement} (section \ref{regpla}). The parallel reconfiguration optimisation reduces the latency of a reconfigurable instruction \textit{miss}, i.e.\ the bitstream has to be loaded via BL1. The register placement dictates the latency of using a reconfigurable instruction that is already programmed inside the instruction disambiguator, which is somewhat analogous to an instruction \textit{hit}, though this latency comes from the fabric logic rather than reading from a cache structure. These are explored together, since they both directly contribute to the behaviour of LUTstructions %
inside the core, as well as the (FPGA) implementation characteristics of the core. 

This exploration is summarised in figure \ref{hmap}, where each colour axis is a softcore implementation metric as reported by Vivado. The brighter the colour appears in the colourmap the more desirable it is in terms of implementation efficiency on the target FPGA device. Each of the 3 plots have the same $x$ and $y$ axes, which are the register placement (S) and configuration parallelism (P). The corresponding secondary 2D axes show how these two variables affect the reconfigurable instruction ``hit'' and ``miss'' latencies respectively.
Between these two latencies, minimising the instruction implementation misses using a high configuration parallelism seems the most challenging. This is because having a stall of hundreds of cycles can %
bottleneck multiprocessing \cite{arc22fpgaext}, while having an instruction latency of tens of cycles is more conventional and can be handled by the design choices relating to ILP. %

Starting from the leftmost plot of figure \ref{hmap}, %
the register utilisation seems unaffected by the register placement optimisation. This is expected, since the implementation always includes the registers in all columns to facilitate the reconfiguration, even with S$>$1. %
This is also the case with conventional FPGA architectures as well, where it is up to the bitstream to involve the registers. When the bitstream loading is complete, then a simple multiplexer arrangement is responsible for bypassing the register on columns with no dedicated registers in all other occasions. 
This dependency is picked correctly by the implementation toolchain, %
and the columns with no register-bypassing circuitry appropriately contribute to the critical path of the soft fabric. This is partly reflected on the rightmost plot of figure \ref{hmap}, where for near S$\ge$8 the fabric starts to reduce the operating frequency of the whole softcore. In terms of logic (LUTs), the register placement optimisation has a certain overhead, because of the multiplexer per LUT4\_4 cell to enable or disable the register for reconfiguration on the columns without register. %

The configuration parallelism affects the resource metrics, but not the %
operating frequency, %
as illustrated in the rightmost plot of figure \ref{hmap}. %
The studied implementation %
accomplishes to keep the critical path unaffected for large degrees of configuration parallelism, even though it also involves the width of the BL1 cache as well. On the other hand, there is the corresponding overhead on FF and LUT usage to achieve the %
bitstream movement demonstrated in section \ref{parrec}. %
The configuration parallelism adds a multiplexer in each LUT4\_4's input whenever its column is the start of the parallel chunks, to be able to inject the bitstream segment. Hence, it contributes to the LUT usage as well. %
Note that the diagonal line that is formed in the colourmap of the middle plot of figure \ref{hmap} relates to how neighbouring multiplexers associated with S and P are co-mapped into logic blocks. 

\subsection{Slot scalability}\label{scala}

A potential scalability concern would be when having a %
higher number of LUTstruction fabrics. %
This is because the instruction disambiguator (ID) needs to multiplex their inputs and demultiplex their outputs efficiently according to the given opcode combination, with some additional complexity for handling the pipeline latency correctly. %
LUTstructions themselves are independent to each other, and are clones of the same logic, so the efficiency reduces to that of the ID.  

The instruction disambiguator is essentially a cache, so the scalability relates to its organisation (number of sets and ways), tag comparison and any other accompanying logic like the replacement policy. The featured implementation is a direct-mapped cache, which is equivalent to 1 way. This minimises the need for a replacement policy, at the cost of potential conflicts and increased reconfigurations. Nevertheless, based on the cost of each
fabric, the efficiency of the approach reduces to the implementation of the caching logic, which is influenced by the rest of the core design. This analogy excludes the memory primitives, as the data (bitstreams) are here stored within the (soft) LUTs of the LUTstruction fabrics. %

Resource-wise, the net overhead per fabric %
is estimated as 11.4K FFs and 15.7K LUTs (%
24\% and 46\% respectively of the unmodified RISC-V softcore \cite{simodense0}). %
This is based on dividing the resources of a 64-slot softcore by 64, under the fixed parameter values of P=16 and  S=7. %
This is while noting the simplifying assumption of not adjusting the rest of the softcore parameters, %
for which further fine-tuning would be desirable in a well-rounded SoC. Notably, the bitstream cache (BL1) size remains steady, %
but a more definitive slot-to-BL1-block ratio could be determined for final SoCs through benchmarking. %

\begin{figure}[h!]
\centering
\includegraphics[width=0.47\textwidth, trim=0 5 -7 5]{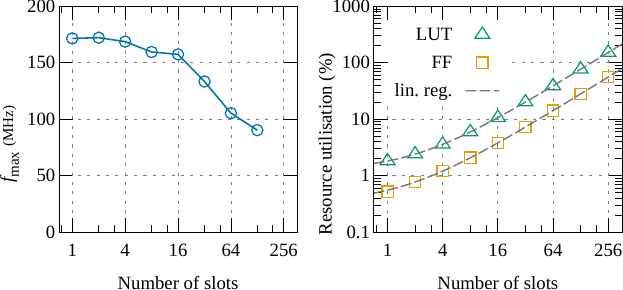}
\caption{Varying the number of LUTstruction fabrics %
on AMD Alveo V80}\label{slots}
\end{figure}

To analyse the scalability of the approach, this multi-slot experiment %
is expanded in figure \ref{slots}. The operating frequency on the left plot initially remains mostly unaffected by the number of fabrics, %
but rapidly decreases after around the 16-slot mark.  The modularity of the design shows favourable behaviour for scalability, as this achieves 128 slots in the same softcore at a respectable 90.3MHz, despite its 77\% LUT utilisation on V80. The right plot of figure \ref{slots} shows the corresponding resource utilisation including the infeasible 256-slot point with 152\% of the LUTs. After overlaying a simple linear regression on the corresponding resources (dashed lines), we observe that the resources scale more or less linearly %
 with the number of slots. Note %
 the perceived wrapping is due to the %
 logarithmic axes and having a steady offset. %

\subsection{System-level performance}\label{slp}

A brief experiment is designed to illustrate that the proposed dynamically\-/loading instructions have a minimal overhead when compared to hard instructions. This seamlessness makes them an attractive acceleration facility for general use. This is studied %
for an artificially-stressful environment, though readers can refer to related work for more insights on the potential acceleration that can broadly be achieved with static FPGA-based custom instructions \cite{ordaz2017making,ordaz2018soft}. %
A loop applies an operation iteratively in the style of STREAM \cite{McCalpin2007}, but it is modified to call either a soft %
instruction, %
or a corresponding software routine for comparison. 

\begin{figure}[h!]
\centering
\includegraphics[width=0.45\textwidth, trim=0 7 -7 5]{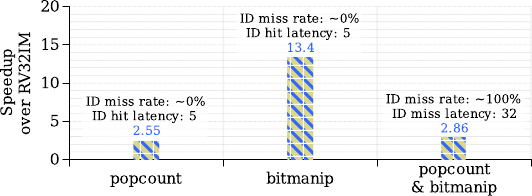}
\caption{End-to-end speedup when %
using FPGAs-on-FPGA as instructions.}\label{pop}
\vspace{-0.1em}
\end{figure}

The first loop variation calls popcount (i.e.\ counts the number of ones) on integers. %
The software implementation is that of GCC's \texttt{\_\_builtin\_popcount()}, where the soft instruction yields a speedup of \emph{2.55x}. %
The second variation applies an arbitrary bitwise permutation per operand and \textit{XOR}s them to represent a more logic-intensive application. The speedup when using the soft instruction over custom software compiled with -O3 is \emph{13.4x}. The final loop variation \textit{XOR}s the results of both of the previous operations. This %
fundamentally
uses two bitstreams in an interleaving fashion.  %
The softcore configuration is purposely reduced to only have one slot  to always cause instruction implementation misses. %
In this extreme case, the high reconfiguration parallelism (P=16, totalling 38.4 GB/s at 150MHz) still manages to yield an overall speedup of \textit{2.86x}. Figure \ref{pop} summarises this experiment. %

\subsection{Out-of-context behaviour}\label{ooc}%

\begin{figure}[h!]
\centering
\includegraphics[width=0.38\textwidth, trim=0 8 0 5]{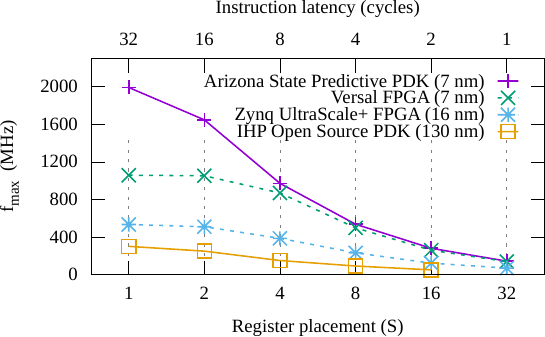}
\caption{%
Operating \(f_{max}\) using different technologies, P=1.}\label{fmax}
\end{figure}

LUTstructions are
also evaluated outside of the softcore %
to quantify the implementation overhead of adding 1 %
instance to a core. %
The FPGA results are produced by Vivado 2025.1.1 for the devices ZU7 and V80, while the silicon-proven 130 nm PDK is used with LibreLane and the ASAP predictive 7 nm PDK %
with  SiliconCompiler. %
Figure \ref{fmax} presents the achievable operating frequencies across the four technologies. %

\begin{figure}[h!]
\centering
\includegraphics[width=0.49\textwidth, trim=0 10 0 6]{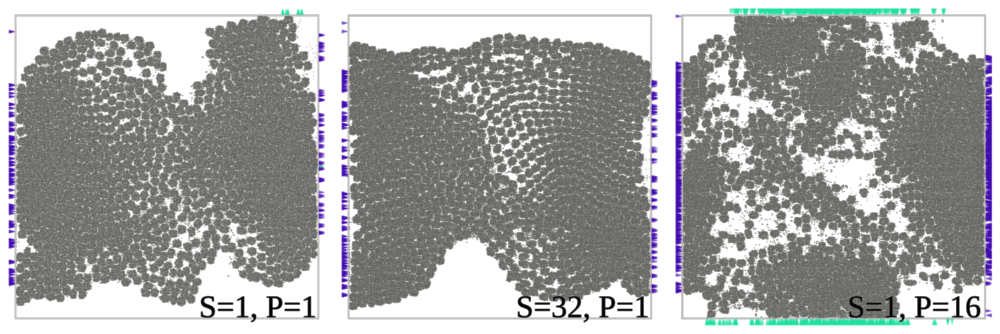}
\caption{ASIC placement images of the fabric at 7 nm.}\label{asap}
\end{figure}

Figure \ref{asap} depicts the placement of a LUTstruction %
into an ASIC for ASAP7. %
The LUT structures are clearly visible forming a 32\(\times\)32 mesh within the chip area. The two leftmost fabrics have a parallelism P of 1, and only differ in the register placement (\(S\)). The second is, hence, more structured, as the length of the critical path is the width of the fabric in LUTs (\(S\)=32). The LUTs in the first area are more ``self-sufficient'' and can be more remote, since their output is always buffered by a register. Alternatively,  the placement could be %
forced into a fixed grid, as with traditional FPGAs. This is not a requirement here, as the pipelined design implies passing universal timing constraints for all possible instructions. %

The rightmost image of figure \ref{asap} shows a variation of the leftmost, but with a high reconfiguration parallelism (P=16). This introduces %
inputs to all edges, as every second column can receive 128 bits per cycle externally totalling 2048 signals. On the FPGA architectures, this has an imperceptible effect on \(f_{max}\) due to the pre-existing routing, %
but on ASICs it has a small overhead %
(i.e.\ 1992 %
to 1694 MHz). %
Nevertheless, all \(f_{max}\) curves of figure \ref{fmax} start from the relative limits of what is achievable on the corresponding technology, such as Vivado's hard limit %
near 1 GHz for the Versal clocking wizard. They also usually remain above ordinary operating speeds, as demonstrated on the rightmost plot of figure \ref{hmap}, where they do not impact the operating frequency of %
the softcore for moderately high \(S\) values.

\section{Discussion}

This exploration answers the \textbf{\emph{research question}} positively. 
After designing and implementing an end-to-end prototype, it is shown that FPGA-based custom instructions can always be %
invoked with a latency of up to the same order of magnitude as AVX instructions \cite{intel}, unlocking their acceleration potential.

This answer also directly addresses the challenges \textbf{\emph{C1}} and \textbf{\emph{C2}} on miss and hit latencies respectively. This is because the working solution exhibits a 32-cycle reconfiguration and a 5-cycle execution latency for custom instructions, with a potential for a wider design space exploration. This is mainly achieved by the fabric's architecture that follows dataflow design patterns (section \ref{desfa}), and by avoiding existing limiting reconfiguration protocols such as ICAP (section \ref{preconf}).

This is still true in the extreme scenario where a %
frequent instruction is always a miss and has to be reprogrammed continuously, while it still achieves speedup over a software-only routine (section \ref{slp}). This follows a worst case approach by %
investigating stressful loops for system-level observations. It is outside of the scope of the paper to consider standard %
benchmark suites, since %
the additional flexibility that comes with FPGA-based instructions %
would drastically shift the focus of the exploration. %
In contrast, prior research like Garp \cite{hauser1997garp} and successor works %
 \cite{goldstein2000piperench, clark2005architecture} provided such a comparison, because they innovated on compilers that combined sets of existing instructions in their coarser-grain fabrics. This dualism in the methodology is consistent throughout the related works mentioned in section \ref{rw}.   %
Therefore, engaging an HLS-style compiler would be appropriate as future work. 
See figure \ref{progr} of section \ref{riscv} on the provided programmability.

Challenge \textbf{\emph{C3}} on operating frequency is rendered insignificant in section \ref{ooc}, where both the FPGA and ASIC-based out-of-context implementations approach the limits of the achievable frequency by the corresponding technology and common designs. This is inline with the design space exploration of section \ref{dses}, since the maximal operating frequency hovers around that of the unaltered core (190 MHz on V80) for most of the latency combinations (figure \ref{hmap}, right), as the fabrics do not shorten the critical path in those instances. The fact that this is achieved as an FPGA-on-FPGA (section \ref{overl}) illustrates the effectiveness of the utilised dataflow techniques. %

Lastly, challenge \textbf{\emph{C4}} on architectural exploration is overcome by the open sourcing of this research, as well as the fact that the hardware is expressed entirely in Verilog. The efficient FPGA-on-FPGA-based fabrics enable the exploration of the architecture independently of vendor-provided technologies and IPs. ICAP, partial reconfiguration and generic eFPGAs are examples of limiting technologies that handicapped related works on reconfigurable instructions (section \ref{rw}), almost all of which do not provide an open source contribution. %

Regarding the architectural details beyond the fabric, we closely follow the baseline softcore \cite{simodense0}. The caches are rather high-end, but %
the core uses a 1-cycle pipeline that simplifies the interaction with functional units. Thus, the modification %
can be abstracted as adding states in a state machine, after specialising a copy of DL1 to work as BL1. %
All code is contributed as a direct description %
of the discussed structures. %

LUTstructions can %
be integrated into other cores since they exhibit simplifying attributes as functional units. By being pipelinable, having a fixed-latency, and operating on the same register file, the integration of the fabrics can be generalised for %
more complex pipelines. In case more slots %
are needed, a scalability study is provided in section \ref{scala}. %
In multi-cores, coherency would be implied, as the configurations are read-only. For further architectural insights on the modelled architecture such as the cache sizes and multiprocessing, see the related feasibility study in \cite{arc22fpgaext}.  %
The presented FPGA architecture is not necessarily definitive, and can be fine-tuned further in future explorations. %
The support of stateful instructions is a desirable next step to harness the already available pipeline registers inside the fabric.

\section{Conclusions}

This paper introduces LUTstructions, %
an efficient open-source FPGA architecture optimised for implementing dynamically-loading reconfigurable instructions. They %
are fast\-/reconfigurable, yield a low instruction latency, and operate at high operating frequencies similar to %
hardened designs. The experimental setup involves their integration in a RISC-V softcore that 
is modified with a bitstream cache to be able to provide the instruction bitstreams on demand with high-throughput for custom instructions of moderate complexity. The reconfiguration is shown to be 28x faster than a more-general state-of-the-art reconfiguration controller. %
An indicative system-level %
benchmark still yields a speedup near 3x over pure software under pressure that causes reconfiguration on every %
LUTstruction call.  
This is entirely achieved on an FPGA, essentially becoming a very high-performing FPGA-on-FPGA, with a 2-to-3 orders of magnitude speedup over related works that are not instruction-optimised. The pipelining and configuration parallelism optimisations achieve similarly-favourable characteristics across different PDK and FPGA technology implementations, making it inviting for wider-adoption and scalability to a high number of slots.
\section*{Acknowledgements}\small

The donation of hardware from AMD AUP and ECS at Southampton is greatly appreciated. No LLM has been used for this research. %

\balance

\bibliographystyle{IEEEtran}
\bibliography{refs}

\end{document}